\documentclass[
 reprint,
 final,
 amsmath,amssymb,
 aps,
 freqn
]{revtex4-2}

\usepackage{graphicx}

\usepackage{dcolumn}
\usepackage{bm}
\usepackage[nodisplayskipstretch]{setspace}
\usepackage{hyperref}

\begin{document}

\preprint{APS/123-QED}
\title{%
 Phonon Feedback Effects on Dynamics of Phase Slip Centers in Finite Gap Superconductors
}

\author{Iris Dorn}
 \email{irdorn@chapman.edu}
 \author{Armen Gulian}
\email{gulian@chapman.edu}
\affiliation{%
 Institute of Quantum Studies, Chapman University\\
 Orange, California, USA
}%

\date{\today}

\begin{abstract}
The results on the behavior of phase-slip centers in thin superconducting wires based on finite-element modeling and time-dependent Ginzburg-Landau (TDGL) equations are discussed. For closer relationship with experiments, we used finite-gap formulation of the TDGL system. Both the dynamic equation for the Cooper-pair condensate $\Psi$-function and the expression for the electric current are more complex than in the gapless case. On this basis, the influence of nonequilibrium phonons is explored. These phonons can essentially change the location of geometrical points in which the phase slippage takes place. They also affect the frequency of phase-slip oscillations. The reported effects are experimentally detectable and can be used in practical devices.
\end{abstract}

\maketitle


\section{Introduction}

    The idea expressed by Bardeen, Cooper and Schrieffer in their microscopic explanation of superconductivity \cite{Bar57},  that the $\Psi$-function of the Ginzburg-Landau theory \cite{Ginz1950}
    at thermodynamic description of superconducting state, may be related to the energy gap in the spectrum of paired electrons was proven by Gor'kov \cite{Gor59}. This proof elevated the phenomenological Ginzburg-Landau equations to the status of microscopic theory: for the BCS-type superconductors, such as NbTiN \cite{Ivry14,hunt1989}, the application of GL-equations is completely adequate to the microscopic approach, at least at temperatures not far from the critical temperature T$_{c}$ of the superconductor. Applicability of the phonon model to objects such as cuprates or iron-based superconductors is still being debated. However, phase-slip centers (PSCs) have been detected in high-temperature cuprate superconductors decades ago \cite{akopyan1990observation}. If the described properties in this report of the phase-slip centers would be found in these materials as well, that will shed additional light onto the current unknowns and help with understanding the puzzling mechanism of superconductivity in these materials.
    
    The generalization of GL-equations for time-dependent problems took considerable effort. The first successful step was by Schmid \cite{schmid1966time}, who came to the conclusion that the proper equation for the $\Psi$-function is not like the Schrödinger's one, but rather has a structure similar to the diffusion equation. Later, this result was confirmed by Éliashberg and Gor'kov \cite{Gor68,gor1970dynamical}, on the basis of the Green's functions approach to superconductivity. Interestingly at that point, a closed system of TDGL equations resulted for gapless superconductors only \cite{abrikosov1960}. After the development of a more powerful energy-integrated Green's functions technique for the kinetic equations in nonequilibrium superconductors, the TDGL set of equations was derived for finite gap superconductors \cite{golub1976,kramer1978,Sch79,Hu80,watts1981}. The first dynamic equation is for the order parameter $\Delta=\left\vert \Delta\right\vert \exp(i\theta)$:
    
\begin{multline}
-\frac{\pi}{8T_{c}}\frac{1}{\sqrt{1+(2\tau_{\varepsilon}\vert \Delta
\vert )^{2}}}\left(  \frac{\partial}{\partial t}+2i\varphi
+2\tau_{\varepsilon}^{2}\frac{\partial\vert \Delta\vert ^{2}}{\partial
t}\right)\Delta  \\
+\frac{\pi}{8T_{c}}\left[  D( \boldsymbol{\nabla}-2i\boldsymbol{A})  ^{2}\right]
\Delta \\ 
+\left[  \frac{T_{c}-T}{T_{c}}-\frac{7\zeta(3)\vert \Delta
\vert ^{2}}{8(\pi T_{c})^{2}}+P(\vert \Delta\vert )\right]
\Delta=0 
\label{1}
\end{multline}

Here the theoretical units
$\hbar$=\textit{c}=\textit{e}=1 are used, \textbf{A} and $\varphi$ are the vector and scalar potentials of
the electromagnetic field, $\tau_{\varepsilon}$ is the electron-phonon relaxation
time, $D$ is the electronic diffusion coefficient, $\zeta(3)$ is the Riemann
zeta function, and $P(\left\vert \Delta\right\vert )$ is the nonequilibrium
phonon term which will be specified below (in absence of the phonon feedback
$P(\vert \Delta\vert )=0$). The order parameter $\Delta$
($\vert \Delta\vert $ is the superconducting energy gap), as was
mentioned above, is proportional to the original Ginzburg-Landau $\Psi
$-function (normalized so as its squared modulus is equal to the density of
pair condensate). Equation (\ref{1}) describes the behavior of the Cooper-pair
condensate, taking into account inelastic electron-phonon collisions. In case
of very strong inelastic collisions ($\tau_{\varepsilon}\rightarrow0$),
Eq.(\ref{1}) converts into its gapless form \cite{Gor68}, where the relaxation of $\Delta$
takes place only due to the condensate itself.

The second equation in the system of GL-equations is for the electric current
density, $\mathbf{j}$. In the static case, the electric field, $\mathbf{E}%
\mathbf{=}\mathbf{-}\mathbf{\dot{A}}\mathbf{-}\mathbf{\nabla}\varphi$,  cannot
exist in superconductors, so that the current contains only the superfluid
motion of pairs: $\mathbf{j}=\mathbf{j}_{s}[\pi\sigma_{n}\left\vert
\Delta\right\vert ^{2}/(4T)]\mathbf{Q}$, where $\mathbf{Q}=\mathbf{-}%
2\mathbf{A}+\nabla\theta.$ In the dynamic case, the current was initially
presented in the form which corresponds to the two-fluid model of
superconductivity \cite{gorter1934}: $\mathbf{j}=\mathbf{j}_{s}+$ $\mathbf{j}_{n},$ where
$\mathbf{j}_{n}=\sigma_{n}\mathbf{E}$ ($\sigma_{n}$ is the conductivity of
normal excitations in the superconductor). Later, it was shown in \cite{gul1987} that,
for the finite gap superconductors, the microscopic theory yields additional
terms in the current and this corresponds to the interference between
superconducting and normal motions of electrons. Thus the current expression
adequate to (\ref{1}) is:

\begin{multline}
\mathbf{j}=\mathbf{j}_{s}+\mathbf{j}_{n}+\mathbf{j}_{int}=\frac{\pi\sigma_{n}
}{4T}\mathbf{Q} \left(  \vert \Delta\vert ^{2}\mathbf{-}
2\tau_{\varepsilon}\frac{\partial\vert \Delta\vert ^{2}}{\partial
t}\right)   \\
+\sigma_{n}\mathbf{E} \Bigg( 1 
+\frac{\vert \Delta\vert
}{2T}\frac{\sqrt{1+(2\tau_{\varepsilon}\vert \Delta\vert )^{2}}}
{2\tau_{\varepsilon}\vert \Delta\vert }  \\ \times \left[  K \left(  \frac
{2\tau_{\varepsilon}\vert \Delta\vert }{\sqrt{1+(2\tau_{\epsilon
}\vert \Delta\vert )^{2}}}\right)  -E\left(  \frac{2\tau_{\epsilon
}\vert \Delta\vert }{\sqrt{1+(2\tau_{\varepsilon}\vert
\Delta\vert )^{2}}} \right) \right] \Bigg) .
\label{2}
\end{multline}
Here $K(x)$ and $E(x)$ are the complete elliptic integrals of the first and
second kind, respectively. Since $[K(x)-E(x)]/x\rightarrow0$ at $x\rightarrow
0$, one can recognize that, as in the case of (\ref{1}) at $\tau_{\varepsilon}=0$, (\ref{2}) yields the gapless limit. With these interference terms
included into the expression for the current, the system of TDGL equations
(\ref{1}) and (\ref{2}) acquires a similar level of accuracy and should be
used for the quantitative description of the dynamics for finite gap superconductors.

The major issue of interest in this article is the phonon feedback, for which
task we need to specify the function $P(\vert \Delta\vert ) $ in
(1). As shown in \cite{gul1987,JJC86}, this function has a form:
\begin{equation}
P(|\Delta|)=-2\tau_{\varepsilon}\operatorname{Re}\int_{0}^{\infty}
d\varepsilon{\frac{\Gamma(\varepsilon)}{\sqrt{(\varepsilon+i\gamma
)^{2}-|\Delta|^{2}}}} ,\label{3}
\end{equation}
where $\gamma=(2\tau_{\varepsilon})^{-1}$. $\Gamma(\varepsilon)$ is related
with the nonequilibrium population of phonons $\delta N_{\omega_{\mathbf{q}}}$
via:
\begin{multline}
\Gamma(\varepsilon)={\frac{\pi\lambda}{2(up_{F})^{2}}}\int_{0}^{\infty
}\omega_{\mathbf{q}}^{2}d\omega_{\mathbf{q}}\int_{|\Delta|}^{\infty
}d\varepsilon^{\prime}\delta(\varepsilon^{\prime} 
+\varepsilon-\omega
_{\mathbf{q}})\\
\times(u_{\varepsilon}u_{\varepsilon^{\prime}}+v_{\varepsilon
}v_{\varepsilon^{\prime}}) (1-n_{\varepsilon}-n_{\varepsilon^{\prime}})\delta
N_{\omega_{\mathbf{q}}}.\label{4}
\end{multline}

In (\ref{4}), $\lambda$ is the dimensionless electron-phonon interaction
constant, $u$\ is the speed of sound in the superconductor, $p_{F}$\ is the
electron Fermi momentum, $\omega_{\mathbf{q}}$ denotes the frequency of a phonon with momentum $\mathbf{q}$, $u_{\varepsilon}=|\varepsilon|\theta(\varepsilon
^{2}-|\Delta|^{2})/\sqrt{\varepsilon^{2}-|\Delta|^{2}}$ is the BCS density of
states for electrons, $v_{\varepsilon}=u_{\varepsilon}|\Delta|/\varepsilon,$
and $n_{\varepsilon}$ is the distribution function of nonequilibrium
electrons. The nonequilibrium phonon distribution function should be found
from the kinetic equation:%
\begin{equation}
{\frac{d}{dt}}(\delta N_{\omega_{\mathbf{q}}})=I(N_{\omega_{\mathbf{q}}%
})+L(N_{\omega_{\mathbf{q}}}),\label{5}%
\end{equation}
where $I(N_{\omega_{\mathbf{q}}})$ is the phonon-electron collision integral,
and $L(N_{\omega_{\mathbf{q}}})$ is the operator describing the phonon exchange of a superconductor with its environment (the heat-bath). In the
simplest approximation \cite{JJC86,Cha77},  the latter may be
defined as: 
\begin{equation}
L(N_{\omega_{q}})\approx-{\frac{\delta N_{\omega_{\mathbf{q}}}}{\tau
_{\mathrm{es}}}}\,,\label{6}%
\end{equation}
where $\tau_{\mathrm{es}} = \alpha d/s$ is the phonon escape time (into the
heat-bath), $d$ is the characteristic dimension of the superconductor (e.g., thickness of the film), $s$ is the speed of sound in it, and $0<\alpha <1$ is a numeric factor typically more than unity \footnote[1]{Chang and Scalapino \cite{Cha77} suggested expression $\alpha$=4/$\eta$, where 4 is a geometrical factor, and $\eta$ is an escape probability (a number between 0 and 1, related with the ratio of acoustic densities of adjacent materials).}. If
$\tau_{\mathrm{es}}\rightarrow0,$ the phonons are in equilibrium, $\delta
N_{\omega_{\mathbf{q}}}\rightarrow0$. However, in many practically important
cases, $\tau_{\mathrm{es}}$\ is finite and (\ref{5}) should be solved
jointly with (\ref{1}) and (\ref{2}). In the so-called  ``generalized
local equilibrium approximation", the solution for $\delta N_{\omega
_{\mathbf{q}}}$ is obtained in \cite{gulian99book,gulian20}. Rather than using this solution,
we will notice that in the absence of population inversion \cite{gulyan1983,lab20},
which until now has not yet been observed experimentally, the integrand in
(\ref{4}), as well as the function $\Gamma(\varepsilon)$ itself, are
positively defined at presence of excess phonons $\delta N_{\omega
_{\mathbf{q}}}>0$. Then, after substitution of $\Gamma(\epsilon)$ into (\ref{3}), we
will find negatively defined function $P(|\Delta|)$. 

In the
first approximation, we will consider this function independent on $|\Delta|$:
$P(|\Delta|)=P$ and simply vary its dimensionless value when solving the TDGL
system to understand the role of this phonon term. As follows from the
analysis of Eq. (\ref{1}), this function effectively reduces  the local value
of critical temperature $T_{c}$. Its non-zero value is caused by external factors, which may have occasional character, like a
film-substrate mismatch or be created deliberately by a heat deposition via
various mechanisms (optical pulses, phonon sources, \textit{etc}.). Thus, during
modeling analysis, we may vary not only its amplitude but also its location in
space and appearance in time.
\section{Dimensionless Equations}
For numerical analysis, it is convenient to represent the system of TDGL
equations in the dimensionless form. For the dimensionless order parameter
$\Psi=\Delta/\Delta_{0}$\ 
(here $\Delta_{0}=\{8\pi^{2}T_{c}(T_{c}%
-T)/[7\zeta(3)]\}^{1/2}$ and $\zeta(3)\approx1.2$\ is the Riemann $\zeta-$
function) one can obtain from (\ref{1}) the following dimensionless equation:%
\begin{multline}
\frac{1}{\sqrt{1+\delta^{2}\vert \Psi\vert ^{2}}}\left(
\frac{\partial}{\partial\tau}+i\text{\ }\bar{\varphi}+\frac{1}{2}\delta
^{2}\frac{\partial\vert \Psi\vert ^{2}}{\partial\tau}\right)
\Psi \\ = 
\xi^{2}(  \boldsymbol{\nabla}-2i\boldsymbol{A})  ^{2}\Psi+(  1-\vert
\Psi\vert ^{2}+p)  \Psi.
\label{7}
\end{multline}
Here, $\delta=2\tau_{\varepsilon}\Delta_{0}$, $\tau=tD/\xi^{2}\equiv t/t_{0}$,
$\bar{\varphi}=2\varphi\xi^{2}/D\equiv2\varphi t_{0}$, $\xi\equiv\xi(T)=\{\pi
D/[8(T_{c}-T)]\}^{1/2}$, where $\xi(T)$ is the ``dirty metal" superconducting
coherence length, and $p=T_{c}/(T_{c}-T)P(|\Delta|)$. The vector potential as
well as the spatial derivatives in (\ref{7}) are not yet dimensionless. The reason is that in the Ginzburg-Landau approach
another spatial parameter, the London penetration length, $\lambda_{L}$, comes
in from the equation for the current density $\mathbf{j}$ (\ref{2}). Using
the relations $\operatorname{curl}\mathbf{H}=4\pi\mathbf{j}%
=\operatorname{curlcurl}\mathbf{A}$, $u=\pi^{4}/[14\zeta(3)]\approx5.798$,
$\eta=(T_{c}-T)/T_{c}\,$, and also choosing the gauge $\varphi=0$, we can
transform (\ref{2}) into:

\begin{multline}
\sigma\dot{A}\Bigg[1+\frac{2}{\pi}\sqrt{\eta u}\vert \Psi\vert
\frac{\sqrt{(\vert \Psi\vert \delta)^{2}+1}}{\vert
\Psi\vert \delta} \bigg[  K\bigg(  \frac{\vert \Psi\vert
\delta}{\sqrt{\vert \Psi\vert ^{2}\delta^{2}+1}}\bigg)  \\ -E\bigg(
\frac{\vert \Psi\vert \delta}{\sqrt{\vert \Psi\vert
^{2}\delta^{2}+1}}\bigg) \bigg] \Bigg]  = 
\\
-\bigg[  A+\frac{i}{2\kappa\vert \Psi\vert ^{2}}(  \psi^{\ast
}\nabla\psi-\psi\nabla\psi^{\ast})  \bigg]\\ \times \bigg(  \vert
\Psi\vert ^{2} 
\mathbf{-}2\delta\sqrt{\frac{\eta}{u}}\frac{\partial
\vert \Psi\vert ^{2}}{\partial\tau}\bigg) 
 -\nabla\times
\nabla\times A
\label{8}
\end{multline}
where $\sigma=\sigma_{n}/t_{0}$ and $\kappa=\lambda_{L}/\xi$ is
the Ginzburg-Landau parameter with the London penetration depth $\lambda
_{L}=\{8\pi^{4}\sigma_{n}(T_{c}-T)/[7\zeta(3)]\}^{-1/2}$. Using this $\lambda
_{L}$ length, we will represent (\ref{7}) in a fully dimensionless form:%

\begin{equation}
\begin{split}
\frac{1}{\sqrt{1+\delta^{2}\left\vert \Psi\right\vert ^{2}}}\frac{\partial
\Psi}{\partial\tau}=-\frac{\delta^{2}}{2\sqrt{1+\delta^{2}\left\vert
\Psi\right\vert ^{2}}}\frac{\partial\left\vert \Psi\right\vert ^{2}}
{\partial\tau}\Psi \\
-\left[\left(\frac{i}{\kappa}\nabla+A\right)\right]^{2}\Psi+\left(  1-\left\vert
\Psi\right\vert ^{2}+p\right)  \Psi
\label{9}
\end{split}
\end{equation}
 in the same $\varphi=0$ gauge as (\ref{8}).

\begin{figure}[!h]
\includegraphics[width=\linewidth]{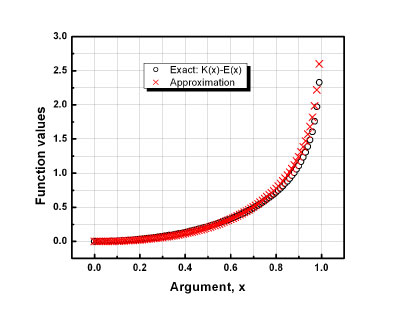}
\caption{Comparison of exact difference of
elliptic functions and its approximation by (\ref{10}).}
\label{Figure1}
\end{figure}

The elliptic integrals in (\ref{8}) can be approximated by elementary functions (Fig. \ref{Figure1}):

\begin{multline}
K(x)-E(x)\cong\frac{\ln(1+x)-\ln(1-x)}{2} \\
+(1-x)\ln(1-x)\equiv\frac{\ln
(1-x^{2})}{2}-x\ln(1-x)\equiv f(x)/x.
\label{10}
\end{multline}

During modeling we used both the exact value and its logarithmic approximation
(\ref{10}) without noticeable difference in solutions.

\section{Results of modeling}
We applied (\ref{8}) and (\ref{9}) to a very thin superconducting wire. In this case the problem is spatially
1-dimensional. Technical details of COMSOL Multiphysics{\textregistered} code are discussed in Appendix A. In brief, we used General Mathematical
Module with three general form PDE interface scalar functions. The complex $\Psi-$function is represented by $\Psi(x) =\operatorname{Re}\Psi(x) +i\operatorname{Im}\Psi(x) $ and  the vector potential is described by a scalar $A(x)$ for the 1-D problem.

The wire is arranged symmetrically at $-L\leq x\leq L$ between the massive
superconducting banks. We accept 
$|\Psi(x=\pm L)|=1$ and $|\Psi^{\prime}(x=\pm L)|=0$, the last of which imposes $\left(  \operatorname{Re}\Psi\right)
^{\prime}(x=\pm L)=0$, $\left(  \operatorname{Im}\Psi\right)  ^{\prime}(x=\pm
L)=0$\ as boundary conditions. For the vector potential we will use the
relation $A(x=\pm L)=-4\pi j_{0}$ which follows from (\ref{8}) and the
boundary conditions for the $\Psi-$function. This external parameter $j_{0}%
$\ regulates the current through the wire: the current is zero when $j_{0}=0$.

\begin{figure}[!h]
\includegraphics[width=\linewidth]{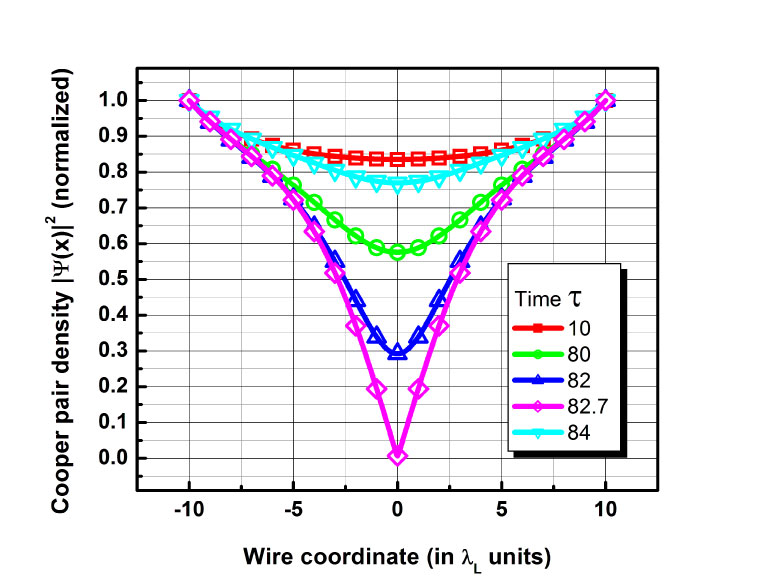}
\caption{Spatial distribution of wave-function modulus at different moments of
time for critical current, $j_{0}=0.4$ and  $\Delta=0.1$ without phonon term, $p=0$.
}
\label{Figure2}
\end{figure}

We will begin with the discussion of the solutions in absence of the phonon
term, $p=0$. When the current exceeds the critical value (in case of Fig. \ref{Figure2}:
$j_{0}=0.4$ for $\Delta=0.1$), the temporal oscillations begin, with $|\Psi|^{2}$
touching zero periodically at the middle of the wire.

\begin{figure}[!h]
\includegraphics[width=\linewidth]{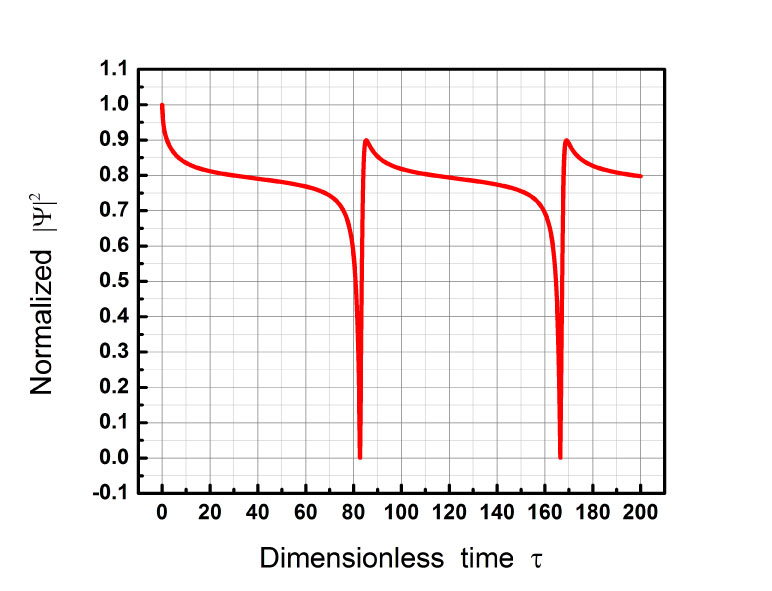}
\caption{Temporal behavior of $|\Psi|$\ at the middle of the wire, $x=0$, for critical current, $j_{0}=0.4$ and  $\Delta=0.1$ without phonon term, $p=0$}
\label{Figure3}
\end{figure}

This behavior of PSCs in 1D wires has been extensively studied in the literature both experimentally and theoretically (to name a few, see \cite{Hu80,watts1981,Vod03,Tia05,xu08,bre12,vodolazov_2007,van12,loz19}, as well as the review articles and books:\cite{Bose_2014, aru08,Bez12,Ivl84,gulian99book,gulian20} and references therein). 

\begin{figure}[!h]
\includegraphics[width=\linewidth]{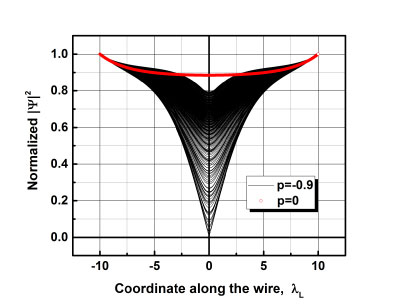}
\caption{At $p=0$ and $j_{0}=0.37$ there are no oscillations; $|\Psi|$\ is
just sagging as shown. At $p=-0.9$ oscillations are present. In both cases
$\Delta=0.1.$}
\label{Figure4}
\end{figure}

\begin{figure}[!h]
\minipage{\linewidth}
\includegraphics[width=0.7\linewidth]{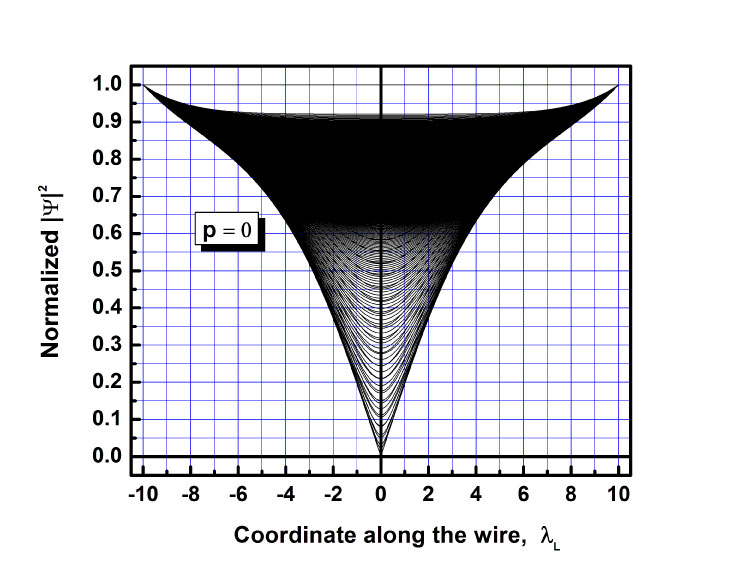}
\endminipage\hfill
\minipage{\linewidth}
\includegraphics[width=0.7\linewidth]{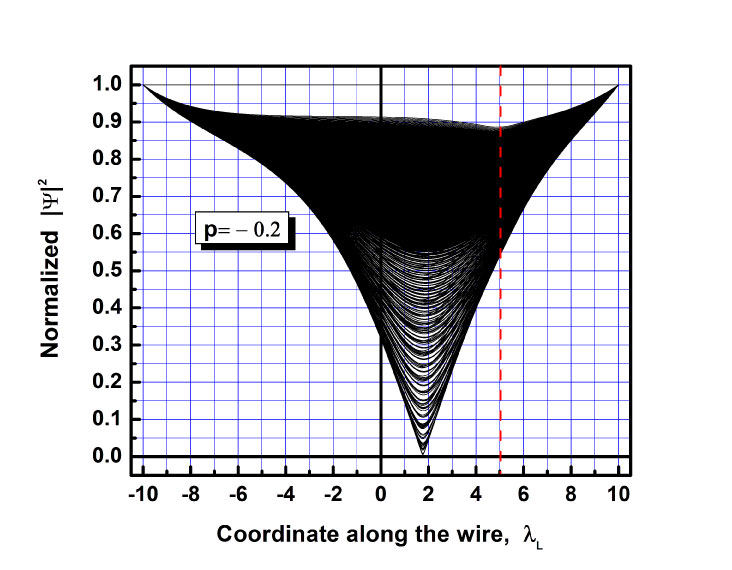}
\endminipage\hfill
\minipage{\linewidth}
\includegraphics[width=0.7\linewidth]{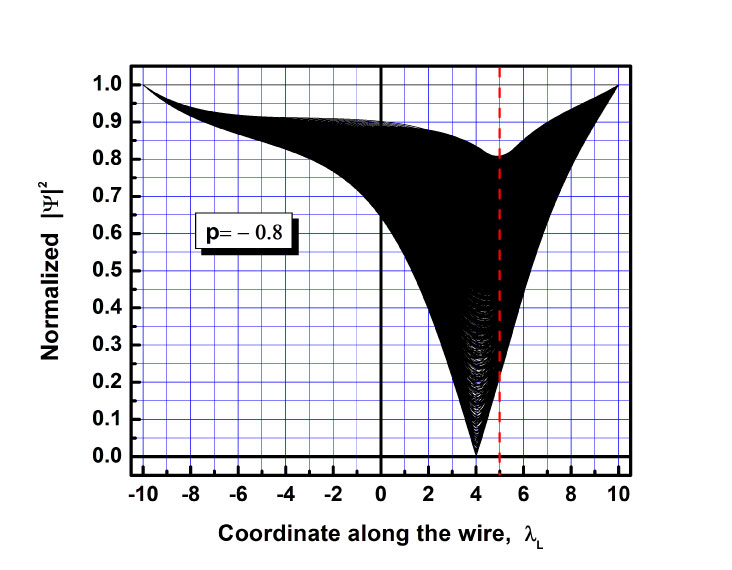}
\endminipage
\caption{Spatial distribution of Cooper pair density for different
intensities of phonon flux action at established periodic PSC oscillations.
$j_{0}$ =0.4 and $\Delta =0.1$. Noticeably, at the beginning of oscillations
period, the sagging starts at the point where the phonon source is located.
At further evolution, the sagging moves towards the geometrical center of
the wire. The hotter the phonon spot is, the closer the location the phase slippage is to it.}
\label{Figure5}
\end{figure}

At this point the phonon term, $p(x,t)$ 
with spatial width $0.1$ (this size is used for all the examples considered further on), 
may be brought in and that will
introduce new features and results. As was explained in Section 1, in case of
positive phonon fluxes (hot spots with elevated temperature along the wire)
$p$ should have negative values. We will start with non-zero $p$ located at
the middle of the wire. Then as described in Fig. \ref{Figure4}, the PSC oscillations may
occur at under-critical values of $j_{0}$.
This is not surprising because in addition to ``weakness"  at the center of the wire caused by the symmetry, the p-term is adding an additional ``weakness" via elevation of ``effective" temperature. It is less obvious what will happen when
the location of the $p-$term is shifted from the central position. To explore, we
located it at $x=5$. Very
interestingly, in this case the actual position at which $|\Psi|^{2}$ is touching
zero is determined in competition of ``geometrical weakness" (\textit{i.e.},
$x=0$) with ``thermal weakness" (\textit{i.e.}, $x=5 $). The higher the $|p|$,
the closer the touching point to the location of $p $. In Fig. \ref{Figure5}, we described
this finding by showing the spatial distribution of temporal variation of
$|\Psi|^{2}$\ for increasingly higher values of $|p|$.

\begin{figure}
\minipage{\linewidth}
\includegraphics[width=\linewidth]{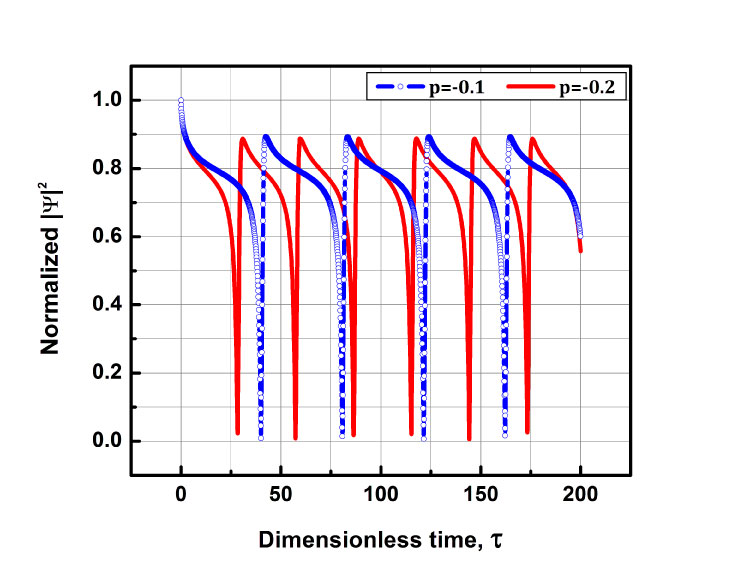}
\endminipage\hfill
\minipage{\linewidth}
\includegraphics[width=\linewidth]{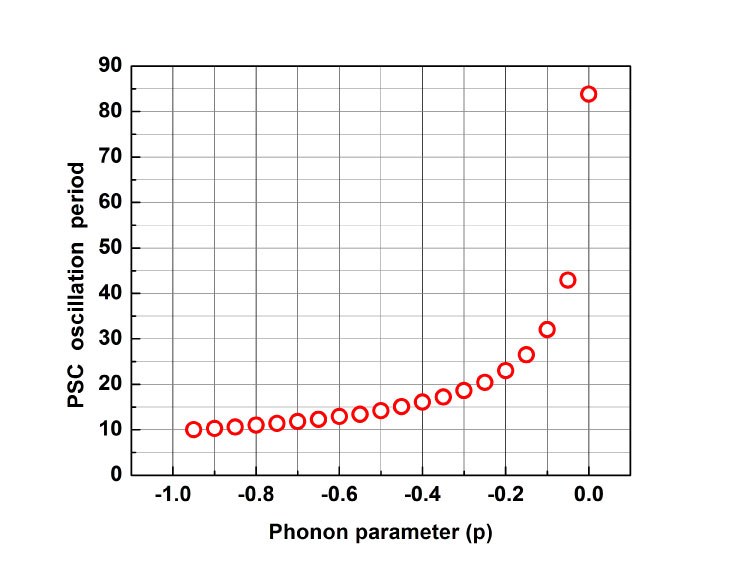}
\endminipage

\caption{Increase of $|p|$ reduces the period of PSC oscillations.}
\label{Figure6}
\end{figure}

Another remarkable feature which the phonon source is introducing is the
dependence of the oscillation frequency on the intensity of the phonon
hot-spot, Fig \ref{Figure6}.

\begin{figure}[h!]
\includegraphics[width=\linewidth]{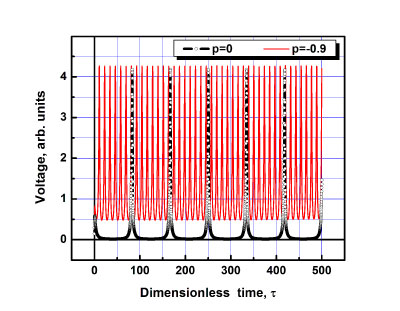}
\caption{Voltage vs. time across the $1D$ wire when PSC\ oscillations take
place with small ($p=0$) and high ($p=-0.9$, located at the center of the
wire) frequencies. Parameters are $\Delta=0.1$ and $j_{c}=0.4$.}
\label{Figure7}
\end{figure}

This observation constitutes the main result of our research. Since the
frequency of oscillations can be measured with high accuracy, this effect may
constitute a basis for high-sensitivity single-particle or radiation
detectors. Indeed, at the absorption of external energy. in the form of photons
or other quanta, the hot-spot is occurring in the substrate which holds the
wire. If the wire is biased by an under-critical current, there are no PSC
oscillations. As soon as the hot-spot is created, oscillations will start with
the frequency proportional to the temperature excursion caused by the event. One of the ways to detect this effect is to measure the voltage between the ends of the wire. Conversion of supercurrent into a normal current and back, which takes
place during the oscillation cycle of PSC, generates the oscillations of the
electric vector $E$ along the wire. Its spatial integral yields oscillatory voltage
$V(t)=%
{\displaystyle\int}
E(t)dl$, which could be measured at the ends of the wire at the constant applied
current $j_{0}$, Fig. \ref{Figure7}.

Though the frequency measurements are the most accurate and possess all the
required information about the hot-spot,
that is not the only possible measurement. A simpler measurement may suffice. The matter is the voltage, as seen from Fig. \ref{Figure7}, is unipolar and its average value is
non-zero. Moreover, at higher frequencies of oscillation the average value is
larger. The average values of voltage for the cases shown in Fig. \ref{Figure7} are
$\widetilde{V}_{high\text{ }frequency}=1.34$ and $\widetilde{V}_{low\text{
}frequency}=0.17$. Thus, the hot spots can be characterized with high accuracy
by a simple DC voltage measurement.

\section{Discussion}

The effects explored above have been associated with non-zero values of the
phonon feedback term $p$. We have considered the simplest case when the
non-zero values of $p$ were caused by external sources such as the hot-spots
induced by a laser or an external phonon flux. A nonequilibrium superconductor
itself provides a more sophisticated origin of the $p-$term. For example, as we
described above, during the process of phase slippage the supercurrent is
transformed into a normal one. Normal current is dissipative and generates
excess phonons. In Section 1, we introduced the phonon escape time,
$\tau_{\mathrm{es}}=\alpha d/u$. If this time is longer than the
phonon-electron relaxation time, then these nonequilibrium phonons may have an
influence on the electrons. This is not the only mechanism. As was shown in \cite{gulyan1986},
breaking and recombination of Cooper pairs creates correspondingly negative
and positive phonon fluxes: $p\varpropto-$ $\partial\left(  |\Psi|^{2}\right)
/\partial t$. Thus, the phase slip centers themselves can serve as a source of
excess phonons. This nonlinear nonequilibrium mechanism
requires additional theoretical investigation which is beyond the scope of this paper.

In previous literature, analogies between Josephson junctions (JJs), or, more
generally, between weakly-coupled superconductors and PSC have been outlined.
Continuing this analogy, we found here that due to the phonon feedback, the
nanowires with phase-slip centers possess features which can be used for
detection purposes. Another aspect of the JJ - PSC analogy is related with the
presence of interference current which we took into account in our
computational scheme. This current is additional to the normal and the superconducting
currents usually used at the TDGL description of PSCs. We used it for a
more proper quantitative description of nonequilibrium effects without
accentuating on its role. However, in the JJ theory there is a paradox of
interference current which may reveal itself in 1D nanowires as well. Future
explorations of the role of the interference current may deliver new discoveries.
\section{Conclusions}
Based on the TDGL equations for the finite gap superconductors, we explored
dynamics of PSCs in 1D nanowires. These equations also contain the term
responsible for the action of nonequilibrium phonons on the order parameter in
the superconductor. Excess nonequilibrium phonons, of the internal or external
origin, are affecting the location of the PSC in the wire. Moreover, they
affect very strongly the frequency of the PSC oscillations. This effect,
discovered and described in detail in our report, may serve as a basis for
practical devices, for example: particle/photon detectors. 
Moreover, it has a deep analogy with exponential increase of PSC oscillation frequency described by Golubev and Zaikin \cite{gol2001} which are in accordance with experimental findings of Bezryadin et al. \cite{bez00}.
The unipolar voltage at
the phase-slip oscillations allows us to suggest a simple DC measurement for
detecting the dynamic effect without necessity of sophisticated measurement
tools for high-frequency signal acquisition. The effect can be used for the PSC
single photon or particle detector (determining the color of the incident photon or the energy of the absorbed particle).
In a more general sense, the phonon feedback may be influential in various
superconducting electronic devices.

\begin{acknowledgments}
This work was supported by the ONR Grants N00014-16-1-2269, N00014-17-1-2972, N00014-18-1-2636, and N00014-19-1-2265.
\end{acknowledgments}

\bigskip
\bigskip

\appendix
\section{APPENDIX}
The \textit{General Form} of \textit{Mathematics Module} of COMSOL 5.5 was
used with the \textit{Time Dependent Study Partial Differential Equation
(PDE)} interface for this simulation. The real and imaginary parts of the
order parameter are represented as $\psi_{1}\equiv\operatorname*{Re}%
(\psi)=u_{1}(x,t)\equiv u$ and $\psi_{1}\equiv\operatorname*{Im}(\psi
)=u_{2}(x,t)\equiv u2$. After separation of real and imaginary parts of
$\Psi-$function, Eq. (\ref{9}) can be represented as:
\begin{multline}
\dot{\psi}_{1}=\frac{1+\delta^{2}\psi_{2}^{2}}{k^{2}\sqrt{1+\delta^{2}
(\psi_{1}^{2}+\psi_{2}^{2})}}(  \psi_{1}{}_{.xx}+\psi_{1}{}_{.yy})\\
-\frac{\delta^{2}\psi_{1}\psi_{2}}{k^{2}\sqrt{1+\delta^{2}(\psi_{1}^{2}
+\psi_{2}^{2})}}(  \psi_{2}{}_{.xx}+\psi_{2}{}_{.yy})\\
+\frac{2(1+\delta^{2}\psi_{2}^{2})}{k\sqrt{1+\delta^{2}(\psi_{1}^{2}+\psi_{2}^{2})}}(A\psi_{2}{}_{.x})\\
+\frac{2\delta^{2}\psi_{1}\psi_{2}}{k\sqrt{1+\delta^{2}(\psi_{1}^{2}+\psi_{2}^{2})}}(  A\psi_{1}{}_{.x}) \\
+\frac{\psi_{2}\sqrt{1+\delta
^{2}(\psi_{1}^{2}+\psi_{2}^{2})}}{k}(  A_{.x})\\  -\frac{\psi_{1}}{\sqrt{1+\delta^{2}(\psi_{1}^{2}+\psi_{2}^{2})}}A^{2}+\frac{\psi_{1}(1-\psi_{1}^{2}-\psi_{2}^{2}-2\mu^{2}+p)  }{\sqrt{1+\delta^{2}(\psi_{1}^{2}+\psi_{2}^{2})}}
\end{multline}
\begin{multline}
\dot{\psi}_{2}=\frac{1+\delta^{2}\psi_{1}^{2}}{k^{2}\sqrt{1+\delta^{2}
(\psi_{1}^{2}+\psi_{2}^{2})}}(  \psi_{2}{}_{.xx}+\psi_{2}{}_{.yy})\\
-\frac{\delta^{2}\psi_{1}\psi_{2}}{k^{2}\sqrt{1+\delta^{2}(\psi_{1}^{2}
+\psi_{2}^{2})}}(  \psi_{1}{}_{.xx}+\psi_{1}{}_{.yy})\\
-\frac{2(1+\delta^{2}\psi_{1}^{2})}{k\sqrt{1+\delta^{2}(\psi_{1}^{2}+\psi_{2}^{2})}}(A_{1}\psi_{1}{}_{.x})\\
-\frac{2\delta^{2}\psi_{1}\psi_{2}}{k\sqrt{1+\delta^{2}(\psi_{1}^{2}+\psi_{2}^{2})}}(A\psi_{2}{}_{.x}) \\
-\frac{\psi_{1}\sqrt{1+\delta^{2}(\psi_{1}^{2}
+\psi_{2}^{2})}}{k}(A_{.x})\\
-\frac{\psi_{2}}{\sqrt{1+\delta^{2}(\psi_{1}^{2}
+\psi_{2}^{2})}}A^{2}+\frac{\psi_{2}(  1-\psi_{1}^{2}-\psi_{2}^{2}
-2\mu^{2}+p)  }{\sqrt{1+\delta^{2}(\psi_{1}^{2}+\psi_{2}^{2})}}
\label{A2}
\end{multline}
The vector potential $A$, Eq. \ref{8}, was represented by $u3(x,t)\equiv u3$. Its equation
has the form:
\begin{multline}
\sigma\dot{A}\Bigg(  1+\frac{2}{\pi}\sqrt{\eta\ast5.798}|\psi|\frac
{\sqrt{(|\psi|\delta)^{2}+1}}{|\psi|\delta}\\ \times \left[  K\left(  \frac{|\psi
|\delta}{\sqrt{|\psi|^{2}\delta^{2}+1}}\right) 
-E\left(  \frac{|\psi|\delta
}{\sqrt{||\psi|^{2}\delta^{2}+1}}\right)  \right]\Bigg)  \\  =-\left[
A+\frac{(  \psi_{2}\nabla\psi_{1}-\psi_{1}\nabla\psi_{2})
}{\kappa|\psi|^{2}}\right]  \\
\times \left(  |\psi|^{2} 
\mathbf{-}2\delta\sqrt
{\frac{\eta}{5.798}}\frac{\partial|\psi|^{2}}{\partial\tau}\right)  -j_{0}
\end{multline}
where $|\psi|=\sqrt{\psi_{1}^{2}+\psi_{2}^{2}}$, and $j_{0}=4\pi\times\left(
current\text{ }density\right)  $.

Partial derivatives $u_{t},u_{x}$and $u_{xx}$ with respect to $t$ and $x$ are
denoted correspondingly as $ut$, $ux$, and $uxx$, and similarly for $u2$ and
$u3$. The form of the PDE to be solved is:
\begin{multline}\label{eq2}
d_{a}=\sigma[  1+(1/\pi)(5.798\cdot\eta(u^{2}+u_{2}^{2}))^{1/2}]\\
\times[exact(1/(1+1/((u^{2}+u_{2}^{2})\delta^{2}))^{1/2})]  ,
\end{multline}
 where $exact$ is an interpolated function corresponding to $f(x)/x$
in (\ref{10}) defined in the set of Global Definitions from an expression for
the Interpolation function based on an imported data table of complete
elliptic integrals $K(x)-E(x)$ (since COMSOL does not have them in its
internal mathematical library). Equivalently for this analytic function, we
also used the logarithmic approximation from Eq. (\ref{10}). The expressions
for terms $\Gamma$ and $F$ for each PDE are given below. For $PDE_{1}$:%
\begin{equation}
\Gamma=-u_{x}/\kappa^{2},
\end{equation}
\begin{multline}
f=-u_{xx}/\kappa^{2}
\\
+\left(  (1+\delta^{2}u_{2}^{2})u_{xx}\right)
/\left(  \kappa^{2}\sqrt{1+\delta^{2}(u^{_{2}}+u_{2}^{2})}\right) \\
-(  \delta^{2}uu_{2}u_{2xx})  /\left(  \kappa^{2}\sqrt{1+\delta
^{2}(u^{2}+u_{2}^{2})}\right) \\
+(  2u_{2x}(1+\delta^{2}u_{2}^{2})u_{3})  /\left(  \kappa
\sqrt{1+\delta^{2}(u^{2}+u_{2}^{2})}\right) \\
+(  2u_{x}\delta^{2}uu_{2}u_{3})  /\left(  \kappa\sqrt{1+\delta
^{2}(u^{2}+u_{2}^{2})}\right) \\
+\left(u_{2}u_{3x}\sqrt{1+\delta^{2}(u^{2}+u_{2}^{2})}\right)/\kappa\\
-uu_{3}^{2}/\sqrt{1+\delta^{2}(u^{2}+u_{2}^{2})}\\
+u\bigg(  1-u^{2}-u_{2}^{2}\\+
p[((x-x_{0})/2)^{2}<width]\bigg)  /\sqrt
{1+\delta^{2}(u^{2}+u_{2}^{2})},
\end{multline}
where $x_{0}$ is the location of the phonon source, and $width$ is its spatial
width. Similarly, for $PDE_{2}$:
\begin{equation}
\Gamma =-u_{2x} /\kappa^{2}
\end{equation}
\begin{multline}
f=-u_{2xx}/\kappa^{2}\\+
(  (1+\delta^{2}u_{2}^{2})u_{2xx})
/\left(  \kappa^{2}\sqrt{1+\delta^{2}(u^{_{2}}+u_{2}^{2})}\right)\\
-(  \delta^{2}uu_{2}u_{xx})  /\left(  \kappa^{2}\sqrt{1+\delta
^{2}(u^{2}+u_{2}^{2})}\right) \\
-(  2u_{x}(1+\delta^{2}u^{2})u_{3})  /\left(  \kappa\sqrt
{1+\delta^{2}(u^{2}+u_{2}^{2})}\right) \\
-(  2u_{2x}\delta^{2}uu_{2}u_{3})  /\left(  \kappa\sqrt
{1+\delta^{2}(u^{2}+u_{2}^{2})}\right) \\
-\left(uu_{3x}\sqrt{1+\delta^{2}(u^{2}+u_{2}^{2})}\right)/\kappa\\
-u_{2}u_{3}^{2}/\sqrt{1+\delta^{2}(u^{2}+u_{2}^{2})}\\
+u_{2}\bigg(  1-u^{2}-u_{2}^{2}\\+
p[((x-x_{0})/2)^{2}<width]\bigg)
/\sqrt{1+\delta^{2}(u^{2}+u_{2}^{2})}.
\end{multline}
For $PDE_{3}$:
\begin{equation}
\Gamma=0\\    
\end{equation}
\begin{multline}
f=-u_{3}[u^{2}+u_{2}^{2}+2\delta a_{11}(2uu_{t}+2u_{2}u_{2t})]\\
+[(uu_{2x}-u_{2}u_{x})/\kappa] \\ \times[1+
2a_{11}(\delta/(u^{2}+u_{2}^{2}
))(2uu_{t}+2u_{2}u_{2t})]-j_{0},
\end{multline}
where $a_{11}=\sqrt{{\eta}/{5.798}}$.
The default Neumann boundary conditions ( $n\cdot\Gamma=0$) for the functions
$\psi_{1}$ and $\psi_{2}$
correspond to the ones specified in Section 3.
 For the vector potential Eq. (\ref{8})
, this equation as well the \textit{Dirichlet boundary condition} is
utilized with the \textit{Prescribed value} $r_{3}=-j_{0}$. In additional,
initial values must be applied for each PDE interface. We used at $t=0$: $u=1
$, $u_{2}=0$ and $u_{3}=-j_{0}$.

At our computations we used
\textit{User-controlled mesh} with \textit{Custom Element Size Parameters}
with \textit{Maximum element size} $0.005$, \textit{Maximum element growth
rate} $1$ and \textit{Resolution of narrow regions} $1$.
\nocite{*}
\providecommand{\noopsort}[1]{}\providecommand{\singleletter}[1]{#1}%

\end{document}